\documentclass[final]{aipproc}

\layoutstyle{6x9}
\newcommand{\bea}{\begin{eqnarray}}
\newcommand{\eea}{\end{eqnarray}}
\newcommand{\eL}{\epsilon_L}
\newcommand{\eLv}{\epsilon_L^{v}}
\newcommand{\eLc}{\epsilon_L^{c}}
\newcommand{\eR}{\epsilon_R}
\newcommand{\eS}{\epsilon_S}
\newcommand{\eT}{\epsilon_T}
\newcommand{\eP}{\epsilon_P}
\newcommand{\teL}{\tilde{\epsilon}_L}
\newcommand{\teR}{\tilde{\epsilon}_R}
\newcommand{\teS}{\tilde{\epsilon}_S}
\newcommand{\teT}{\tilde{\epsilon}_T}
\newcommand{\teP}{\tilde{\epsilon}_P}

\begin{document}

\begin{flushright}
NPAC-12-13
\end{flushright}
\vspace{-1.0cm}

\title{$\beta$ decays in the LHC era:\\ from ultracold neutrons to colliders}

\classification{%<Replace this text with PACS numbers; choose from this list:                \texttt{http://www.aip..org/pacs/index.html}>
}

\author{Mart\'in Gonz\'alez-Alonso\footnote{This work was supported by the U.S. DOE contract DE-FG02-08ER41531 and the Wisconsin Alumni Research Foundation. I would like to thank the CIPANP 2012 organizers for the nice conference and my various collaborators in this field for the pleasure of working with them.}}
{address={Theoretical Nuclear, Particle, Astrophysics, and Cosmology (NPAC) Group,\\ Department of Physics, University of Wisconsin-Madison, WI 53706, USA}}

\begin{abstract}
In this talk I review the New Physics reach of semileptonic beta decay experiments, and their interplay with LHC searches. Assuming the new particles are heavy enough we can use an Effective Field Theory (EFT) approach to analyze the LHC searches, what allows us to perform a direct and model-independent comparison with low-energy experiments.
\end{abstract}

\maketitle

Semileptonic beta decays, that played such an important role in the development of the Electroweak Theory, represent a very useful tool in the current search of New Physics (NP). We review here the implications of the current and future experiments with neutrons, nuclei and mesons, that will achieve an unprecedented level of precision, as nicely shown in several talks at this conference \cite{exp-talks}. We also review the interplay with LHC searches in the $pp\to e \bar{\nu}$ channel, that probe the same kind of NP since the underlying partonic process is the same ($\bar{u}d\to e\bar{\nu}$).

NP bounds extracted from nuclear and neutron beta are usually expressed using the $C$ coefficients of the effective hadronic Lagrangian, where protons and neutrons are the active fields \cite{Jackson1957zz, Severijns2006dr}. Likewise, the bounds from pion decays are expressed in terms of the low-energy constants of the effective $\chi PT$ Lagrangian. However, all these beta decays are probing the same partonic process and hence it is worth going one step down in the theoretical description and using the quark-level Lagrangian, so that all the experimental results can be casted in a common language and one can compare them and evaluate their interplay. For this reason we will work with the following low-scale  $O(1 \ {\rm GeV})$  effective Lagrangian for semi-leptonic transitions:
\bea
{\cal L}_{\rm CC}  &=&
- \frac{G_F V_{ud}}{\sqrt{2}} \  \Big[ \ \Big( 1 +  \eL \Big) \  
\bar{e}  \gamma_\mu  (1 - \gamma_5)   \nu_{\ell}  \cdot \bar{u}   \gamma^\mu  (1 - \gamma_5)  d   \\
&+& \teL  \ \ \bar{e}  \gamma_\mu  (1 + \gamma_5)   \nu_{\ell}  \cdot \bar{u}   \gamma^\mu  (1 - \gamma_5)  d  \nonumber\\
&+&   \eR   \  \   \bar{e}  \gamma_\mu  (1 - \gamma_5)   \nu_{\ell}  \cdot \bar{u}   \gamma^\mu  (1 + \gamma_5)  d  
\ + \  \tilde{ \epsilon}_R   \  \   \bar{e}  \gamma_\mu  (1 +  \gamma_5)   \nu_{\ell} \cdot \bar{u}   \gamma^\mu  (1 + \gamma_5)  d  \nonumber\\
&+&  \eT   \   \bar{e}   \sigma_{\mu \nu} (1 - \gamma_5) \nu_{\ell}    \cdot  \bar{u}   \sigma^{\mu \nu} (1 - \gamma_5) d
\ + \  \teT      \   \bar{e}   \sigma_{\mu \nu} (1 + \gamma_5) \nu_{\ell}    \cdot  \bar{u}   \sigma^{\mu \nu} (1 + \gamma_5) d  \nonumber \\
&+&  \eS  \  \  \bar{e}  (1 - \gamma_5) \nu_{\ell}  \cdot  \bar{u} d  \ + \  \teS  \  \  \bar{e}  (1 +  \gamma_5) \nu_{\ell}  \cdot  \bar{u} d  \nonumber \\
&-& \eP  \  \   \bar{e}  (1 - \gamma_5) \nu_{\ell}  \cdot  \bar{u} \gamma_5 d 
\ - \  \teP  \  \   \bar{e}  (1 + \gamma_5) \nu_{\ell}  \cdot  \bar{u} \gamma_5 d  \ \Big]+{\rm h.c.}~. \nonumber
\eea
The non-standard effective couplings $\epsilon_i$ and $\tilde{\epsilon}_i$ are functions of the masses and couplings of the new heavy particles yet to be discovered, in the same way as the Fermi constant is a function of the W mass and the weak coupling. However we do not need to know these functions to compare the NP reach of different beta decay experiments.   

This step from hadrons to quarks requires the calculation of the associated form factors, usually a non-trivial task due to the non-perturbative character of QCD at low energies. Schematically we can write $C = \rm{FF} \times \epsilon$, where $C$ are the hadronic level coefficients, and FF are the form factors. Thus we can see that the extraction of the NP couplings $\epsilon$ is a very rich physics problem that involves: (i) experimental determination of the different observables; (ii) analytical methods that connect the observables with the hadronic coefficients $C$, including subleading effects like weak-magnetism or radiative corrections; and (iii) numerical or analytical methods to calculate the form factors. 

\begin{figure}%[!b]
\centering
\hspace{-0.1in}
\includegraphics[width=0.45\textwidth]{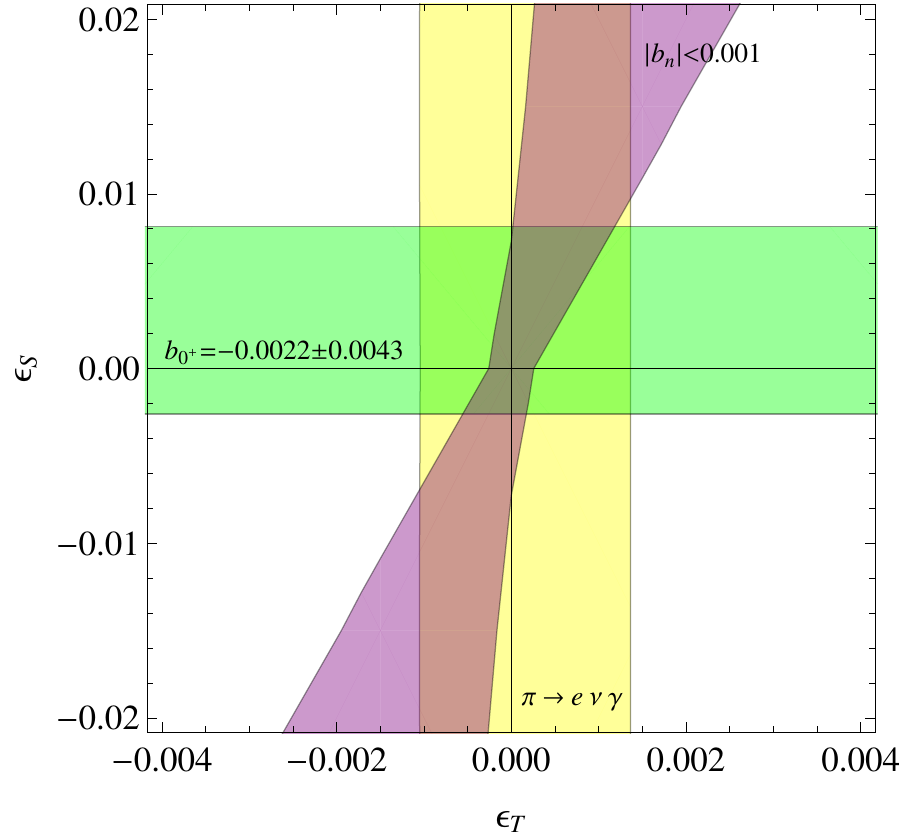}
\hspace{0.2in}
\includegraphics[width=0.45\textwidth]{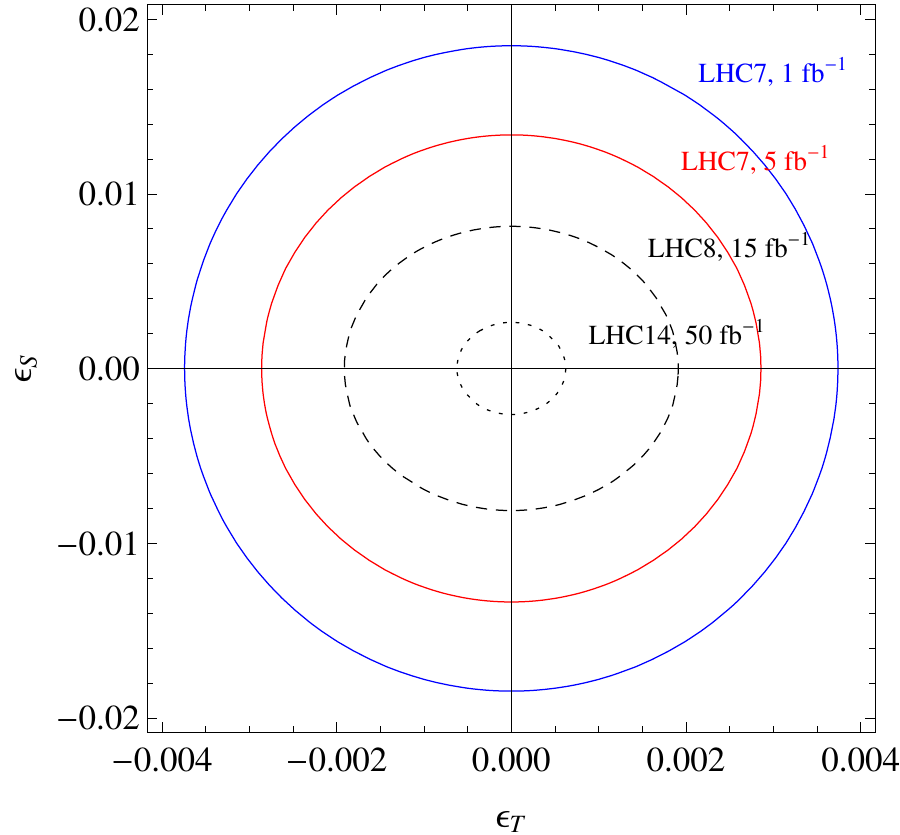}
\caption{90\% C.L. limits on the scalar and tensor NP couplings $\epsilon_{S,T}$. %$\eS$ and $\eT$.
 The left figure shows the bounds from superallowed nuclear decays \cite{Hardy2004id} (green) and radiative pion decay \cite{Bychkov:2008ws} (yellow), along with the expected bound from future measurements of the Fierz term $b$ in neutron decay %\cite{Pocanic:2008pu,UCNb} 
\cite{exp-talks} 
 (purple). The hadronic form factors have been taken from Refs.~\cite{Bhattacharya:2011qm,Mateu:2007tr}. 
The right figure shows the bounds obtained with 5 fb$^{-1}$ of data recorded at $\sqrt{s}=$ 7 TeV by the CMS Coll. in the $pp\to e\nu$ channel \cite{ENcms5fb} (blue solid line). The previous bound obtained using only $1~\rm{fb}^{-1}$ of data \cite{Bhattacharya:2011qm} is shown for comparison (red solid line), as well as the expected bound with higher luminosities and energies (dashed and dotted lines). We refer the reader to Refs.~\cite{Bhattacharya:2011qm,VCMGMGA} for more details.}
\label{fig:ST}
\end{figure}

%These ten parameters $\epsilon_i$ and $\epsilon_i^\prime$, that encode all the NP impact on beta decays, have a very different phenomenology. 
The $\epsilon_i^\prime$ coefficients affect the observables only quadratically, since they come with a right-handed (RH) neutrino, whereas the $\epsilon_i$ terms can interfere with the Standard Model (SM), and thus beta decay experiments are more sensitive to them. The pseudo-scalar and (axial)vector couplings are strongly constrained by the ratio $\rm{R_\pi = \Gamma(\pi\to e\nu) /\Gamma(\pi\to \mu\nu)}$ \cite{Britton:1992pg} and CKM unitarity tests \cite{Cirigliano:2009wk} respectively, whereas the situation for the scalar and tensor couplings $\eS$ and $\eT$ is summarized in Fig.~\ref{fig:ST} (left)\footnote{Although not shown in Fig.~\ref{fig:ST}, the ratio $R_\pi$ is also a very powerful probe of scalar and tensor interactions since they generate radiatively a pseudo-scalar interaction \cite{Campbell:2003ir}. See Refs.~\cite{Bhattacharya:2011qm,VCMGMGA} for more details.}. A thorough review of the bounds on these coefficients from nuclear, neutron and meson decays can be found in Refs.~\cite{Bhattacharya:2011qm,VCMGMGA}. 

As mentioned above, one expects that whatever NP is generating non-zero $\epsilon_i$ coefficients should as well modify the $pp\to e\nu$ cross-section at the LHC at some level. Moreover, if we assume that the new particles are too heavy to be produced at the LHC we can keep using a model-independent EFT description\footnote{The matching conditions between Wilson coefficients of the low- and high-energy effective Lagrangians can be found in Refs.~\cite{Bhattacharya:2011qm,VCMGMGA}.} and we can put bounds on the $\epsilon_i$ coefficients from the LHC searches. In this scenario, the cross-section with transverse mass higher than $\overline{m}_T$ takes the following form\footnote{Notice that high-energy searches probe separately the vertex correction $\eLv$ and contact $\eLc$ contributions to the coupling $\eL$, defined in Ref.~\cite{VCMGMGA}.}:
\bea
\label{eq:sigmamt}
\sigma_{pp\to e\nu}(m_T \!\!>\! \overline{m}_T) &=&    
\sigma_W \Big[ ( 1 +   \eLv)^2+  |\teL|^2  + | \eR|^2 \Big] - 2 \, \sigma_{WL}\,  \eLc \left( 1 +  \eLv \right) \\
&&\hspace{-2.1cm}
+~ \sigma_R  \Big[ |\teR|^2 \!+ |\eLc|^2 \Big] + \sigma_S \Big[ |\eS|^2  \!+  |\teS|^2  \!+ |\eP|^2 \!+  |\teP|^2   \Big] +  \sigma_T  \Big[ |\eT|^2  \!+  |\teT|^2     \Big]~,\nonumber 
\eea 
%Here $\sigma_W (\bar{m}_T)$ is the SM contribution, so we can see that the LHC is not very sensitive to $\eLv$, $\teL$ and $\eR$, whereas $\sigma_{WL, R, S,T}(\bar{m}_T)$ are several orders of magnitudes larger than the SM contribution, which explicit form can be found in Ref.~\cite{REF}. 
where $\sigma_W (\bar{m}_T)$ is the SM contribution and $\sigma_{WL, R, S,T}(\bar{m}_T)$ are new functions, which explicit form can be found in Ref.~\cite{VCMGMGA}. All we need to know is that they are several orders of magnitudes larger than the SM contribution, what compensates for the smallness of the NP coefficients and makes possible to set significant bounds on them from these searches. On the other hand, we see that the LHC is not very sensitive to $\eLv$, $\teL$ and $\eR$.

The absence of any significant excess of high-$m_T$ events in this channel can be translated into bounds on the different NP couplings, as shown in Fig.~\ref{fig:ST} (right) for scalar and tensor couplings. From Eq.~\eqref{eq:sigmamt} it is clear that the bounds for $\epsilon_{S,P}$ and $\tilde{\epsilon}_{S,P}$ are the same, and likewise for $\eT$ and $\teT$. It can be shown that the sensitivity to the remaining two couplings $\teR$ and $\eLc$ is approximately a factor two weaker than in the $\eT$ case \cite{VCMGMGA}.

As Fig.~\ref{fig:ST} nicely shows, there is an interesting competition between low- and high-energy searches looking for new scalar and tensor interactions. For other couplings, like the pseudoscalar ones, low-energy probes are much more powerful, whereas for interactions involving RH neutrinos the LHC will dominate the search \cite{VCMGMGA}. We see that only the combination of both searches can give us a complete picture of non-standard charged current interactions.

\bibliographystyle{aipproc}   % if natbib is available
%\bibliography{Vincenzo-modified,Michael-modified,Martin-modified}

\end{document}